\documentstyle[12pt,fleqn]{article}
\begin{document}

\title{Supermembranes in 4 Dimensions}
\author{Jens Hoppe\\[1cm]Institut fr Theoretische Physik\\
 Universitt Karlsruhe\\P.O. Box 6380\\76128 Karlsruhe\\Germany\\
 e--mail: be10@dkauni2.bitnet\\[2cm]KA--THEP--9--93}
\date{October 93}
\maketitle
\vspace{4cm}
\begin{abstract}
\begin{normalsize}
A non--parametric gauge for supermembranes is introduced which
substantially simplifies previous treatments and directly leads
to the supersymmetric extension of a K\'{a}rm\'{a}n--Tsien gas.
\end{normalsize}
\end{abstract}
\thispagestyle{empty}

\newpage
As shown in \cite{1}, the dynamics of a relativistic membrane in 4
space--time dimensions is closely related to that of a 2--dimensional
irrotational isentropic gas, whose pressure--density relation if of
the `K\'{a}rm\'{a}n--Tsien' form \cite{2}, i.e.~with the pressure being
(up to a constant) inversely proportional to minus the mass density. While
certain special properties of such a pressure--density relation have
long been known (see e.g.~\cite{3}), another one is that by adding
two real anticommuting fields to the velocity--potential and the
mass--density, the theory can be extended to a supersymmetric one.
While the corresponding Hamiltonian was written down in \cite{1},
it's detailled derivation \cite{4} --- which was somewhat lengthy when
starting with the partially gauge fixed light cone supermembrane
Hamiltonian (as given in \cite{5}) --- can actually be simplified
substantially by using a non--parametric description of supermembranes
from the start. This description seems to be the simplest one yet known.
It consists of choosing `right at the beginning' 3 of the 4 Minkowski
lightcone coordinates as the parameters of the world--volume.

In this gauge, the supermembrane--Lagrangean (\cite{6}; respectively
\cite{5}, eq.~(2.1)) simplyfies enormously, becoming
\begin{equation}
{\cal L}=-\sqrt{2(\dot{p}-\frac{i}{2}\Theta^{tr}\dot{\Theta})+
 (\vec{\nabla}p-\frac{i}{2}\Theta^{tr}\vec{\nabla}\Theta)^2\,}+
 \frac{i}{2}\Theta^{tr}\vec{\gamma}\times\vec{\nabla}\Theta
\end{equation}
where $p=p(t,x,y)$ is a (real) bosonic field (the light--cone
coordinate $x^0+x^3$, as a function of the other three, i.e.~%
$t=(x^0-x^3)/2,x=x^1,y=x^2), \Theta=\Theta(t,x,y)=(\Theta_1,
\Theta_2)^{tr}$ real fermionic, and $ \vec{\gamma}\times\vec{\nabla} =
 \left(\begin{array}{cc}0&1\\1&0\end{array}\right)\partial_y -
 \left(\begin{array}{cc}1&0\\0&-1\end{array}\right)\partial_x\;.$\\
Defining canonical momenta,
\begin{eqnarray}
&&\tilde{q}:=\frac{\delta{\cal L}}{\delta\dot{p}}=
 \frac{-1}{\sqrt{2(\dot{p}-\frac{i}{2}\Theta^{tr}\dot{\Theta})
 +(\vec{\nabla}p-\frac{i}{2}\Theta^{tr}\vec{\nabla}\Theta)^2\,}\:}\\
&&\Pi_r:=\frac{\delta^L{\cal L}}{\delta\dot{\Theta_r}}=
 -\frac{i}{2}\,\frac{\Theta_r}{\sqrt{2(\dot{p}-\frac{i}{2}
 \Theta^{tr}\dot{\Theta})+(\vec{\nabla}p-\frac{i}{2}\Theta^{tr}
 \vec{\nabla}\Theta)^2\,}\:}=
 +\frac{i}{2}\Theta_r\tilde{q}\nonumber
\end{eqnarray}
one finds
\begin{equation}
\dot{p}\tilde{q}+\dot{\Theta_1}\Pi_1+\dot{\Theta_2}\Pi_2-{\cal L}=
 -\frac{1}{2}\tilde{q}\left\{(\vec{\nabla}p-\frac{i}{2}\Theta^{tr}
 \vec{\nabla}\Theta)^2+\frac{1}{\tilde{q}^2}+\frac{1}{\tilde{q}}\,
 \frac{i}{2}\,\Theta^{tr}\vec{\gamma}\times\vec{\nabla}\Theta\right\}.
\end{equation}
Letting
\begin{equation}
\Psi:=\frac{\Theta_1+i\Theta_2}{\sqrt{2}}\,\sqrt{q},\quad q:=-\tilde{q},
\end{equation}
one obtains the Hamiltonian $(\partial=\partial_x-i\partial_y)$
\begin{equation}
H=\frac{1}{2}\int dx\,dy\left\{q(\vec{\nabla}p-\frac{i}{2}\,
 \frac{\Psi\vec{\nabla}\bar{\Psi}+\bar{\Psi}\vec{\nabla}\Psi}{q})^2+
 \frac{1}{q}(1+i(\Psi\bar{\partial}\Psi+\bar{\Psi}\partial\bar{\Psi}))\right\},
\end{equation}
where the classical Dirac--brackets to be used are
\begin{equation}
\left[q(t,\vec{x}),\,p(t,\vec{x}')\right]=\delta(\vec{x}-\vec{x}'),\quad
 \left[\Psi(t,\vec{x}),\,\bar{\Psi}(t,\vec{x}')\right]=
 -i\delta(\vec{x}-\vec{x}')
\end{equation}
(all others=0). The expression for the remaining 9 generators of the
Poincar\'{e} group follow in analogy with the purely bosonic case (cp.~%
\cite{1}). Also, one may easily verify that
\begin{equation}
Q=\int d\vec{x}\left(\frac{\Psi}{\sqrt{q}}+\sqrt{q}\bar{\Psi}\partial p+
 \frac{i}{2}\,\frac{\Psi\bar{\Psi}}{\sqrt{q}}\,\partial\bar{\Psi}\right)
\end{equation}
and its complex conjugate, $\bar{Q}$, satisfy \cite{4}
\begin{equation}
[Q,Q]=0,\quad[Q,\bar{Q}]=-2iH
\end{equation}

\section*{Acknowledgement}
I would like to thank M. Bordemann and O. Lechtenfeld for very helpful
discussions.


\begin{thebibliography}{9}
\bibitem{1} M. Bordemann, J. Hoppe; {\sl Phys.~Lett.~B 317} (1993)
\bibitem{2} H. S. Tsien; {\sl J. of Aeron.~Sciences 6} (1939) 399\\
            T. von K\'{a}rm\'{a}n; {\sl J. of Aeron.~Sciences 8} (1941) 377
\bibitem{3} N. Coburn; Quat.~J. of Appl.~Math.~(1945) 106\\
            C. Rogers, W. F. Shadwick; `Bcklund Transformations
            and their Applications', {\sl Academic Press} 1982
\bibitem{4} J. Hoppe; KA--THEP--6--93
\bibitem{5} B. de Wit, J. Hoppe, H. Nicolai; {\sl Nuc.~Phys.~B 305} (1988) 545
\bibitem{6} E. Bergshoeff, E. Sezgin, P.K. Townsend; {\sl Phys.~Lett.~B 189}
            (1987)
\end{thebibliography}
\end{document}